\documentclass[twocolumn]{aastex63}
\usepackage{amsmath}
\usepackage{longtable}
\usepackage{bm}
\usepackage{mathrsfs}
\usepackage[normalem]{ulem}
\usepackage{layouts}
\usepackage[caption=false]{subfig}

\newcommand{\lya}{Ly$\alpha\,$}

\newcommand{\cii}{[\ion{C}{2}]}

\revised{\today}
\submitjournal{ApJ}

\shorttitle{The effective lifetime of the quasar population at $z\sim6$}
\shortauthors{Morey et al.}
\graphicspath{{./}{figures/}}

\begin{document}

\title{Estimating the effective lifetime of the $z\sim6$ quasar population from the\\ composite proximity zone profile}

\correspondingauthor{Karna Morey}
\email{kmorey@mit.edu}

\author[0000-0003-4768-3155]{Karna A. Morey}
\affiliation{MIT Kavli Institute for Astrophysics and Space Research, 77 Massachusetts Ave., Cambridge, MA 02139, USA}

\author[0000-0003-2895-6218]{Anna-Christina Eilers}\thanks{NASA Hubble Fellow}
\affiliation{MIT Kavli Institute for Astrophysics and Space Research, 77 Massachusetts Ave., Cambridge, MA 02139, USA}

\author[0000-0003-0821-3644]{Frederick B. Davies}
\affiliation{Lawrence Berkeley National Laboratory, CA 94720-8139, USA}
\affiliation{Max-Planck-Institute for Astronomy, K\"onigstuhl 17, 69117 Heidelberg, Germany}

\author[0000-0002-7054-4332]{Joseph F. Hennawi}
\affiliation{Physics Department, University of California, Santa Barbara, CA 93106-9530, USA}

\author[0000-0003-3769-9559]{Robert A. Simcoe}
\affiliation{MIT Kavli Institute for Astrophysics and Space Research, 77 Massachusetts Ave., Cambridge, MA 02139, USA}

\begin{abstract}
The lifetime of quasars can be estimated by means of their proximity zone sizes, which are regions of enhanced flux bluewards of the Lyman-$\alpha$ emission line observed in the rest-frame UV spectra of high-redshift quasars, 
because the intergalactic gas has a finite response time to the quasars' radiation. We estimate the effective lifetime of the high-redshift quasar population from the composite transmitted flux profile within the proximity zone region of a sample of $15$ quasars at $5.8\leq z\leq 6.6$ with precise systemic redshifts, and similar luminosities, i.e.\ $-27.6\leq M_{1450}\leq-26.4$, and thus a similar instantaneous ionizing power. 
We develop a Bayesian method to infer the effective lifetime from the composite spectrum, including robust estimates of various sources of uncertainty on the spectrum. 
We estimate an effective lifetime of the quasar population as a whole of $\log_{10}(t_{\rm Q}/{\rm yr}) = 5.7^{+0.5 (+0.8)}_{-0.3 (-0.5)}$ given by the median and $68$th ($95$th) percentile of the posterior probability distribution. 
While our result is consistent with previous quasar lifetime studies, it poses significant challenges on the current model for the growth of supermassive black holes (SMBHs) located in the center of the quasars' host galaxies, which requires that quasar lifetimes are more than an order of magnitude longer. 
\end{abstract}

\keywords{Supermassive black holes --- 
Quasars--- High-redshift galaxies---Early universe --- method: Astronomy data analysis}

\section{Introduction} \label{sec:intro}

Quasars are powered by accretion onto a central supermassive black hole (SMBH), additionally, they are the most luminous, non--transient objects in the known universe and thus they can be detected and observed at very early cosmic epochs
\citep[e.g.][]{Mortlock2011A7.085, Venemans2013DiscoverySurvey, Wu2015An6.30,Mazzucchelli20176.5, Banados2018An7.5, Onoue20196.7, Yang2020Poniuaena:Hole, Wang2020AJ02520503}. High--redshift quasars at $z\gtrsim 6$ are of particular interest for studying the early formation and growth of SMBHs. Observations show that their SMBHs have masses on the order of $\gtrsim 1$ billion solar masses already at a time when the universe is still very young, i.e. $\lesssim 1$ Gyr after the Big Bang. 

These black holes are believed to grow from an initial seed via accretion during luminous quasar phases. The duration of the luminous quasar phase, known as the quasar lifetime $t_Q$, is an important parameter to understand the formation and growth of these SMBHs \citep{Inayoshi2020TheHoles}. Assuming
exponential growth, i.e. 
\begin{align}
    M_{\rm BH}(t_Q)=M_{\rm seed}\exp\left(\frac{t_Q}{t_S}\right)
\end{align}
with an e-folding or Salpeter time 
\begin{align}
    t_S\simeq4.5\times 10^7\left(\frac{\epsilon}{0.1}\right)\left(\frac{L_{\rm bol}}{ \bm{L}_{\rm edd}}\right)^{-1}~\rm yr
\end{align}
it requires at least 0.8 Gyr to grow a $10^9$ $M_{\odot}$ SMBH from a 100 $M_{\odot}$ seed with a fiducial radiative efficiency of 10\% \citep{Shakura1973BlackAppearance} and accretion at the Eddington limit, i.e. $L_{\rm bol} = L_{\rm edd}$. It remains an open question as to the validity of this simple exponential growth model
and the possibility of additional physics leading to more complicated light curves  \citep[e.g.][]{DiMatteo2005EnergyGalaxies, Springel2005SimulationsQuasars,Hopkins2005LuminositydependentFunction,Novak2011FeedbackModels, Davies2019b}.

At high redshifts, the age of the universe is comparable to the timescale required for SMBHs to form \citep{Volonteri2010FormationHoles}. However, measurements of the timescales of quasar activity have proven to be difficult. At lower redshifts of $z\sim2-4$ previous studies have estimated the duty cycle of quasars, which denotes the total fraction of time compared to the age of the universe that galaxies spent as luminous quasars and therefore represents an upper limit on the quasars' lifetime $t_Q$, by comparing the number density of quasars to the abundance of dark matter halos via clustering studies
\citep{Efstathiou1988High-redshiftCosmogony,Haiman2001ConstrainingClustering,Martini2001QuasarQuasars, Martini2004CoevolutionGalaxies}. These studies rely on the measurement of the abundance of quasar host dark matter halos in comparison to the abundance of luminous quasars to estimate the quasars' duty cycle \citep{White2008ConstraintsClustering}.
However, these methods have yielded only weakly constrained estimates of $\sim 10^6 - 10^9$ years, since clustering estimates are susceptible to high uncertainty due to variance in parameters governing how quasars populate dark matter halos \citep{Shen2009QuasarProperties, White2012TheSurvey, Conroy2013ADemographics,Cen2015AQuasars}. Other estimates of quasar activity timescales have been made using an extension of the 
``Soltan argument'' \citep{Soltan1982MassesQuasars}, where the quasar luminosity function can be compared to quiescent SMBHs observed in galaxies locally, which have led to an estimate of the quasars' duty cycle to between $10^7-10^8$ years \citep{Yu2002ObservationalHoles}. 

Furthermore, previous studies have used the quasars' light echo transverse to the line--of--sight to measure the time delay between the onset of the quasar emission and changes in opacity of the intergalactic medium \citep{Adelberger2004MeasuringEffect, Hennawi2007QuasarsQuasars,Schmidt2017StatisticalYears, Schmidt2018ModelingObscuration, Bosman20205.8} to constrain the lifetime of quasars to be $t_Q\sim10^5-10^7$ years. 

Recent studies have shown how the lifetime of quasars $t_Q$ can be inferred by means of the proximity zone region observed in rest-frame UV and optical spectra of high redshift quasars \citep{Khrykin2016TheQuasars,Eilers2017a, Eilers2018b, Khrykin2019EvidenceQuasars, Davies2019a, Eilers2020}. The proximity zone is a region of enhanced transmitted flux in the vicinity of quasars that has been ionized by the quasar's own radiation 
 \citep{Bajtlik1988QuasarRedshift, Haiman2001ProbingReionization, Wyithe20056, Bolton2007AMedium, Lidz2007QuasarReionization, Bolton2011HowJ1120+0641, Keating2015ProbingQSOs}. Estimating quasar lifetimes from the HI proximity zone is a technique that has primarily been used to estimate $t_Q$ of high redshift quasar populations, but recent studies have used the Helium II line at $z \simeq 3-4$ to measure individual quasar lifetimes and the quasar lifetime distribution \citep{Khrykin2021,Worseck2021DatingEffect,Khrykin2019EvidenceQuasars}. The intergalactic medium has a finite response time to the quasar's radiation and therefore the level of enhanced flux due to ionized gas around the quasar is highly sensitive to the lifetime of quasars. 

\cite{Eilers2017a, Eilers2018b} used proximity zones in the \ion{H}{1} \lya forest to estimate the lifetime for $z \simeq 6$ quasars.  The same method can be applied to proximity zones in the \ion{He}{2} \lya forest to estimate the lifetimes of quasars at low redshift ($z \simeq 3-4$), as was done in \cite{Khrykin2019EvidenceQuasars}.
%
\cite{Eilers2017a} found a population of quasars $z \simeq 6$ that show very small proximity zones and therefore are likely to be very young objects ($t_Q \lesssim10^5 \mathrm{~yr}$). 

Assuming a simple ``light-bulb'' light curve for the quasars, i.e. the quasar turns on abruptly and emits at a constant luminosity during its entire lifetime $t_Q$, the probability of detecting a quasar age of $t_Q^{(i)}$ is $p = t_Q^{(i)}/ t_Q$. Given that these young quasars corresponded to $\simeq10\%$ of the population, this suggested a fiducial lifetime of $t_Q = 10^6$~yr for the quasar population \citep{Eilers2017a, Eilers2020}. \cite{Davies2019b} expanded upon this work by developing a semi-analytic model of quasar proximity zones, including non-equilibrium interactions between the IGM and the quasar, using hydrodynamical radiative transfer simulations to simulate the proximity zone behavior on different timescales. The observed distribution of proximity zones by \cite{Eilers2017a} was similar to the modeled distribution for $t_Q \simeq 10^6$ using the ``lightbulb'' model. As part of this analysis, \cite{Davies2019b} considered more general light curves, not just the ``lightbulb'' model, and found consistent results for the lifetime, again confirming evidence for $t_Q \simeq 10^6$~yr. 


\cite{Davies2019a} measured the lifetime of individual  high redshift quasars during the epoch of reionization (EoR), and measured the age of these quasars to be roughly $\simeq10^6$~yr. By using the neutral IGM as a counter of the ionizing photons emitted by the quasar, they were able to constrain the lifetimes of two $z\gtrsim7$ quasars. By measuring the number of emitted ionizing photons, they were able to directly measure the radiative efficiency $\epsilon \simeq 0.09\%$, much smaller than the canonical 10\% value given by Eddington limited accretion. This study notes, however, that if this behavior is instead caused by UV obscured quasars, these radiative efficiencies are consistent with a fraction of UV obscured quasars greater than $82\%$ at redshift $z\simeq 7$. However, further work is needed to characterize whether this mechanism is indeed radiative inefficiency or UV obscuration. 


These lifetime estimates are more than an order of magnitude shorter than expected and cause tension with the current model for the growth of SMBHs. A potential solution could be the presence of very massive initial black hole seeds (i.e. $M_{\text{seed}}>1000M_{\odot}$), vastly exceeding the mass of Population III stellar remnants \citep{Lodato2006SupermassiveDiscs, Visbal2014DirectHaloes, Habouzit2016OnSeeds,Schauer2017TheVelocities}. 
As mentioned above, very low radiative efficiencies could also explain such short lifetimes, with radiative efficiency rate $\epsilon \lesssim 0.01 - 0.001$ previously suggested by other studies \citep{Volonteri2015TheRedshift, Trakhtenbrot2017OnQuasars, Davies2019a}, or highly UV obscured black hole growth phases \citep{Eilers2018b, Davies2019a}. 
Additional estimates of $t_Q$, both for individual quasars and for the entire population, could provide important constraints on these mechanisms, i.e. provide constraints on the mass of the initial seeds or the specific radiative efficiency rate $\epsilon$.

In this study, we will estimate the effective lifetime of the high-redshift quasar population by means of the \textit{composite} transmitted flux within the proximity zone region. To this end, we assembled a data set of $15$ quasar spectra at $z\sim 6$ with a similar absolute magnitude, i.e. a similar ionizing radiation output. By comparing this stacked proximity zone profile to outputs from radiative transfer simulations of quasars at different lifetimes, we constrain the effective lifetime of the quasar population. 

This paper is organized as follows: in \S~\ref{sec:data} we discuss our quasar sample and the selection criteria. In \S~\ref{sec:methods} we describe our quasar continuum estimation procedure and the radiative transfer models. In \S~\ref{sec:results}, we discuss our stacking procedure, as well as the forward modeling of uncertainties, and present our final result. We summarize our results in \S~\ref{sec:summary}.

Throughout this paper, we assume a flat $\Lambda$CDM cosmology of $h = 0.685$, $\Omega_m = 0.3$, and $\Omega_{\Lambda} = 0.7$, which is consistent within the $1\sigma$ errorbars from \citet{PlanckCollaboration2020PlanckParameters}. 

\begin{deluxetable*}{cllcccccc}[!t]
\tablecaption{Overview of our data sample. \label{tab:quasars}}
\setlength{\tabcolsep}{6pt}
\rotate
\tablehead{
\colhead{Name} & \colhead{RA [hms]} & \colhead{DEC [dms]}& \colhead{Redshift} &\colhead{$\mathrm{M}_{1450}$} & \colhead{Telescope/Instrument}  & \colhead{SNR} &\colhead{PID} &\colhead{PI}}
    \startdata
     PSOJ011+09	&00:45:33.566 &	+09:01:56.93& 6.4693  &-26.85 &	VLT/X-Shooter & 7 & 0101.B-0272(A) & Eilers\\
PSOJ036+03& 02:26:01.873 &	+03:02:59.24&	6.5412	&-27.28	&VLT/X-Shooter& 11& 0102.A-0154(A)/	0100.A-0625(A) & D'Odorico\\
PSOJ065-26& 04:21:38.050&	-26:57:15.72&	6.1877&	-27.25&	VLT/X-Shooter & 19&098.B-0537(A) & Farina\\
SDSSJ0842+1218&08:42:29.438	&+12:18:50.47&	6.0763&	-26.91 &VLT/X-Shooter & 26&097.B-1070(A)& Farina\\
ATLAS/PSOJ158-14& 10:34:46.509 &	-14:25:15.86&	6.0681&	-27.41&VLT/X-Shooter & 15&096.A-0418(B)& Shanks\\
PSOJ159-02&	 10:36:54.190 &	-02:32:37.940&6.3809&	-26.59	&VLT/X-Shooter & 9&098.B-0537(A)&Farina\\
SDSSJ1143+3803 &11:43:38.347 &	+38:08:28.823	&5.8367&	-26.69 &	Keck/DEIMOS&15 & 2017A\_U078 & Hennawi\\
SDSSJ1148+5251&11:48:16.64 &	+52:51:50.2&	6.4189&	-26.81 &	Keck/ESI&37 & U43E, U11E, U2E & Becker\\
&&&&&&& C70E & Djorgovski\\
&&&&&&& H25aE, H45aE & Cowie\\
&&&&&&& C97E & Kulkarni\\
SDSSJ1306+0356& 13:06:08.259 & +03:56:26.19&	6.0337&	-26.81&	VLT/X-Shooter & 38&084.A-0390(A)&Ryan-Weber\\
ULASJ1319+0950& 13:19:11.291 &+09:50:51.49&	6.1330&	-26.88&	VLT/X-Shooter&52&084.A-0390(A)&Ryan-Weber\\
SDSSJ1335+3533& 13:35:50.81 &	+35:33:15.82&	5.9012&	-26.67 & Keck/DEIMOS&22 & 2017A\_U078 & Hennawi\\
PSOJ217-16&14:28:21.371	&-16:02:43.73&	6.1498	&-26.93 &	Magellan/FIRE& 9 && Simcoe\\
CFHQSJ1509-1749&	15:09:41.781&-17:49:26.68&6.1225&	-27.14&	VLT/X-Shooter&37&085.A-0299(A)&D'Odorico\\
PSOJ323+12& 21:32:33.178 &	+12:17:55.07&6.5872&	-27.06 & VLT/X-Shooter&14&098.B-0537(A)&Farina\\
PSOJ359-06&23:56:32.452	&-06:22:59.31&	6.1719&	-26.79&	VLT/X-Shooter&19&	098.B-0537(A)&Farina\\
\enddata
    \tablecomments{The sub-mm redshift estimate for PSOJ036+03 was obtained by \cite{Venemans2015ALMA}, the redshift estimate for J1319+0950 was obtained by \cite{Wang2013StarALMA}, the redshift estimate for SDSSJ1335+3533 was obtained by \cite{Wang2013StarALMA}, the redshift estimates for SDSSJ1143+3808, PSOJ359-06, PSOJ158-14, and PSOJ011+09 were obtained by \cite{Eilers2020}, and all others were published in \cite{Decarli20185.94}. The SNR values are estimated using a pixel scale of 0.05 {\AA}/pixel in the rest frame of the quasars.} 
\end{deluxetable*}

\section{Data Sample} \label{sec:data}

For our analysis we select quasars at $5.8 < z < 6.6$, and an absolute magnitude at 1450 {\AA} in the rest-frame within a narrow range of $-26.6 \lesssim M_{1450}\lesssim -27.4$. We require that all objects in our sample have available medium-resolution 
spectroscopy at optical and possibly also near-infrared (NIR) wavelengths.
Additionally, we select quasars for which precise redshift estimates ($\Delta v\lesssim 100\,\rm km\,s^{-1}$) based on sub-mm observations of the \cii\ emission line at $158\,\mu$m ($\nu_{\rm rest}=1900.548$~GHz) or the CO(6--5) emission line at $3\,\rm mm$ ($\nu_{\rm rest} = 691.473\rm\,GHz$) from the Atacama Large Millimetre Array (ALMA) or the NOrthern Extended Millimeter Array (NOEMA) at the Institute de Radioastronomie Millim{\'e}trique (IRAM) are available, since uncertainties on the quasars' systemic redshift contribute the largest source of uncertainty to the proximity zone measurements \citep{Eilers2017a, Eilers2020}. 
Furthermore, we remove all quasars from the sample that show broad absorption line features (BALs) or proximate damped \lya\ systems (pDLAs), which would cause additional absorption in the proximity zone region and would bias our results. These selection criteria result in $15$ objects with a mean redshift of $\langle z\rangle =6.23$, and mean magnitude of $\langle M_{1450}\rangle =-26.9$. An overview of our data sample is shown in Table \ref{tab:quasars}. 

Of the quasars in our sample, eleven were observed with 
the X-Shooter instrument at the Very Large Telescope (VLT) \citep{Vernet2011X-shooterTelescope}, two where observed using the DEep Imaging Multi-Object Spectrograph (DEIMOS) instrument at the Keck observatory \cite{Faber2003}, one was observed using the Echellette Spectrograph and Imager (ESI) instrument also at the Keck observatory \citep{Sheinis2002ESIImager}, and one was observed using the Folded-port InfraRed Echellette (FIRE) spectrometer instrument at the Magellan telescope \citep{Simcoe2013FIRE:Telescopes}. The individual detector resolutions $R=\frac{\lambda}{\Delta \lambda}$ are $R\approx8800$ for the X-Shooter instrument \cite{Vernet2011X-shooterTelescope}, $R\approx6000$ for the FIRE instrument\citep{Simcoe2013FIRE:Telescopes}, $R\approx5000$  for the DEIMOS instrument \citep{Faber2003}, and $R\approx5400$ for the ESI instrument \citep{Sheinis2002ESIImager}.


\subsection{Data Reduction}

All optical and NIR spectroscopic data observed with VLT/X-Shooter and Keck/DEIMOS were reduced applying the open source python spectroscopic data reduction package \texttt{PypeIt} \citep{Prochaska2020PypeIt:Pipeline}.
For the reduction we first subtract the sky emission, which was performed on the 2D images by including both image differencing between dithered exposures whenever these were available, and a B-spline fitting procedure \citep[e.g.][]{Bochanski2009MASE:Spectrograph}. 
We apply an optimal spectrum extraction technique \citep{Horne1986AnSpectroscopy} to extract the 1D spectra. All individual 1D spectra are flux calibrated using the standard stars $\rm LTT\,3218$ (for spectra observed with VLT/X-Shooter) and $\rm G191B2B$ or $\rm Feige\,110$ (for spectra taken with Keck/DEIMOS). These fluxed 1D spectra are stacked and a telluric model and the quasar PCA model from \cite{Davies2018} are jointly fitted to the stacked spectra using telluric model grids from the Line-By-Line Radiative Transfer Model \citep[LBLRTM\footnote{\url{http://rtweb.aer.com/lblrtm.html}};][]{Clough2005AtmosphericCodes, Gullikson2014CORRECTINGCODE}. 

\begin{figure*}[!t]
    \centering
    \includegraphics[width = \textwidth]{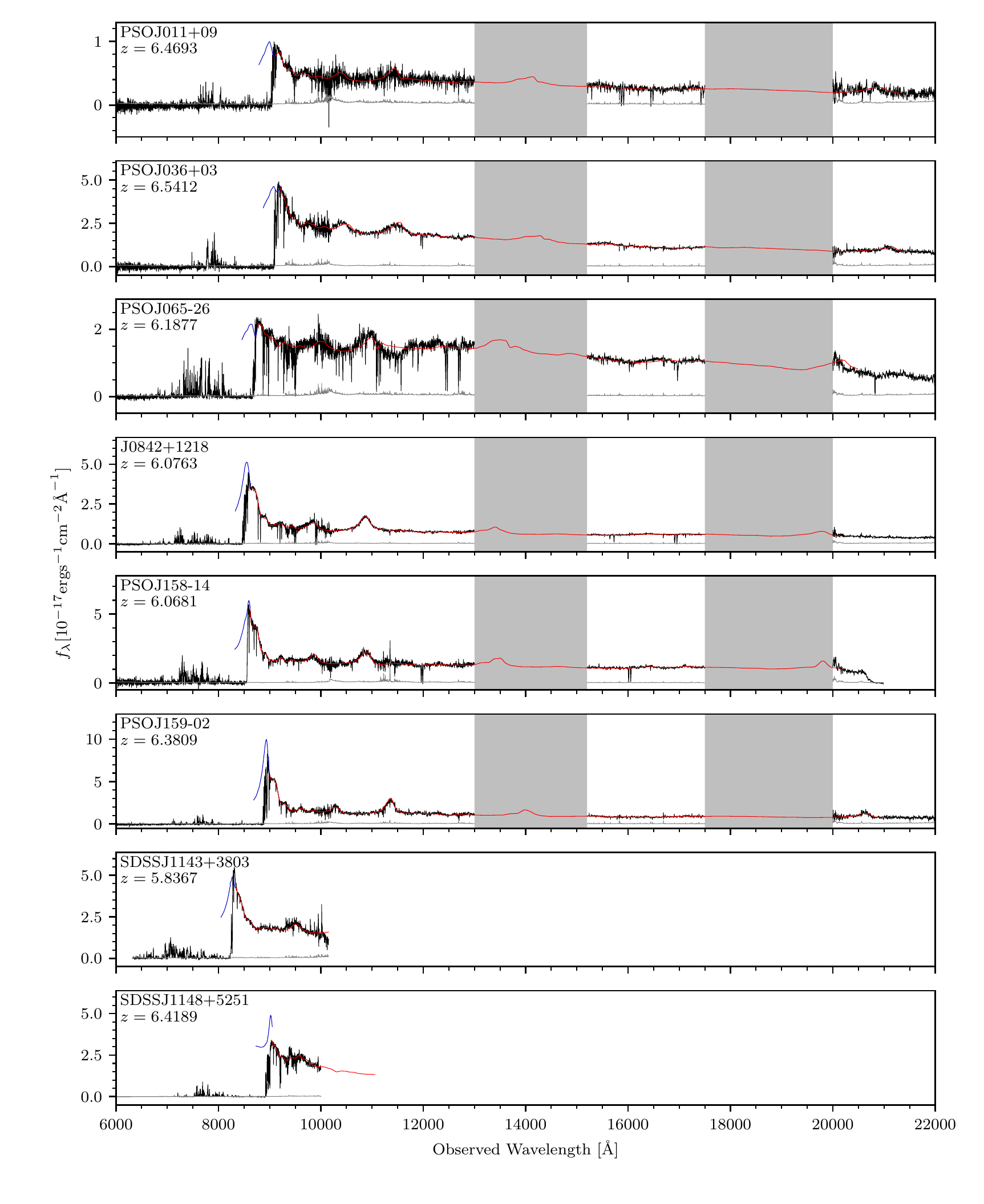}
    \caption{\textbf{Quasar spectra from our sample.} We mask the regions of telluric absorption in grey. All fluxes (black) and noise vectors (grey) have been inverse-variance smoothed with a 10 pixel filter. We also show the quasar continuum estimates fitted on the red-side (red) and the blue-side prediction (blue) from our PCA analysis described in section \ref{sec:cont}.}
    \label{fig:spectra1}
\end{figure*}

\begin{figure*}[!t]
    \centering
    \includegraphics[width = \textwidth]{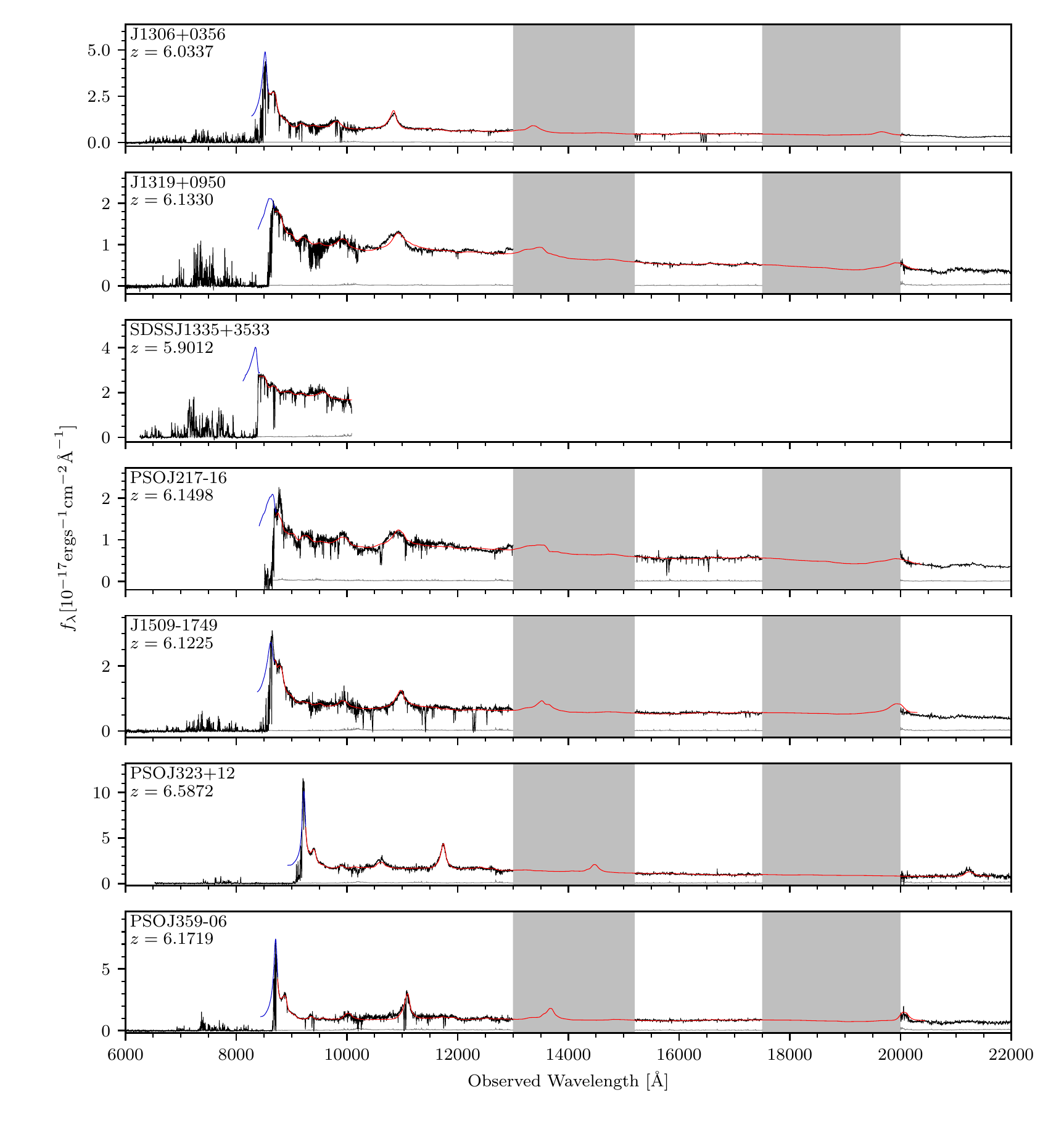}
    \caption{Same as Fig. \ref{fig:spectra1} for the remaining objects in our sample. }
    \label{fig:spectra2}
\end{figure*}

We then match the visible (VIS) and NIR regions for the VLT/X-Shooter spectra. To this end, we compute the mean value of all pixels between 10070 and 10270 {\AA} in the observed frame, and then scale the NIR spectra according to a constant factor such that the two means are equal. In this step, we do not normalize the spectra to unity, but merely scale the NIR spectra to be matched with the VIS spectra.

For the FIRE spectra, which do not require matching between NIR and visible parts, we use the IDL-based
pipeline \texttt{FIREHOSE}\footnote{\url{http://web.mit.edu/\~rsimcoe/www/FIRE/ob_data.htm}}, developed by \citet{Simcoe2010}. The procedures used for data reduction include cosmic-ray rejection, flat-fielding, wavelength calibration using OH sky lines (calibrated to vacuum wavelength), nonlinearity correction, optimal extraction of one-dimensional spectra \cite{Horne1986AnSpectroscopy}, combining individual exposures, merging multiple echelle orders, and heliocentric corrections. This pipeline performs sky subtraction via a exposure-specific B-spline model of the sky, based on the technique developed in \cite{Bochanski2009MASE:Spectrograph}. 
We reduce the ESI spectra using the \texttt{ESIRedux} pipeline\footnote{\url{http://www2.keck.hawaii.edu/inst/esi/ESIRedux/}} developed as part of the XIDL\footnote{\url{http://www.ucolick.org/~xavier/IDL//}} suite of astronomical routines in the Interactive Data Language (IDL). This pipeline is very similar to \texttt{FIREHOSE}. More details about modifications to this pipeline for high-redshift quasars can be found in \cite{Eilers2017a}.

In the end we mask all spectral regions that are affected by telluric absorption, namely between 13000 and 15200 {\AA}, 17500 and 20000 {\AA}, and beyond 22500 {\AA} in the observed frame. Figures~\ref{fig:spectra1} and \ref{fig:spectra2} show the final spectra of all $15$ quasars in our sample. 
We estimate the SNR of the spectra by taking the median pixel-based ratio of the signal to noise between 1265 and 1295 {\AA} in the rest frame of the quasar, due to the absence of any emission lines in this region. We scale the pixel-based SNR for each of the instruments assuming Gaussian white noise to the X-shooter resolution to compare the SNR measurements with a common resolution in Table \ref{tab:quasars}.


\section{Methods} \label{sec:methods}

We aim to estimate the lifetime of the quasar population as a whole
by means of their composite proximity zone region. Thus, in order to obtain the stacked transmission profile within the quasars' proximity zones, we first have to normalize the spectra by their continuum emission. Once the quasars are continuum normalized, we stack them between 1190 and 1218 {\AA} in the rest--frame, in order to obtain an estimate for the composite flux transmission profile in the proximity zone region. After stacking the quasars, we compare this stacked flux profile to 1-d radiative transfer models. 

\subsection{Quasar Continuum Normalization}\label{sec:cont}

\begin{figure*}
    \centering
    \includegraphics[width = \textwidth]{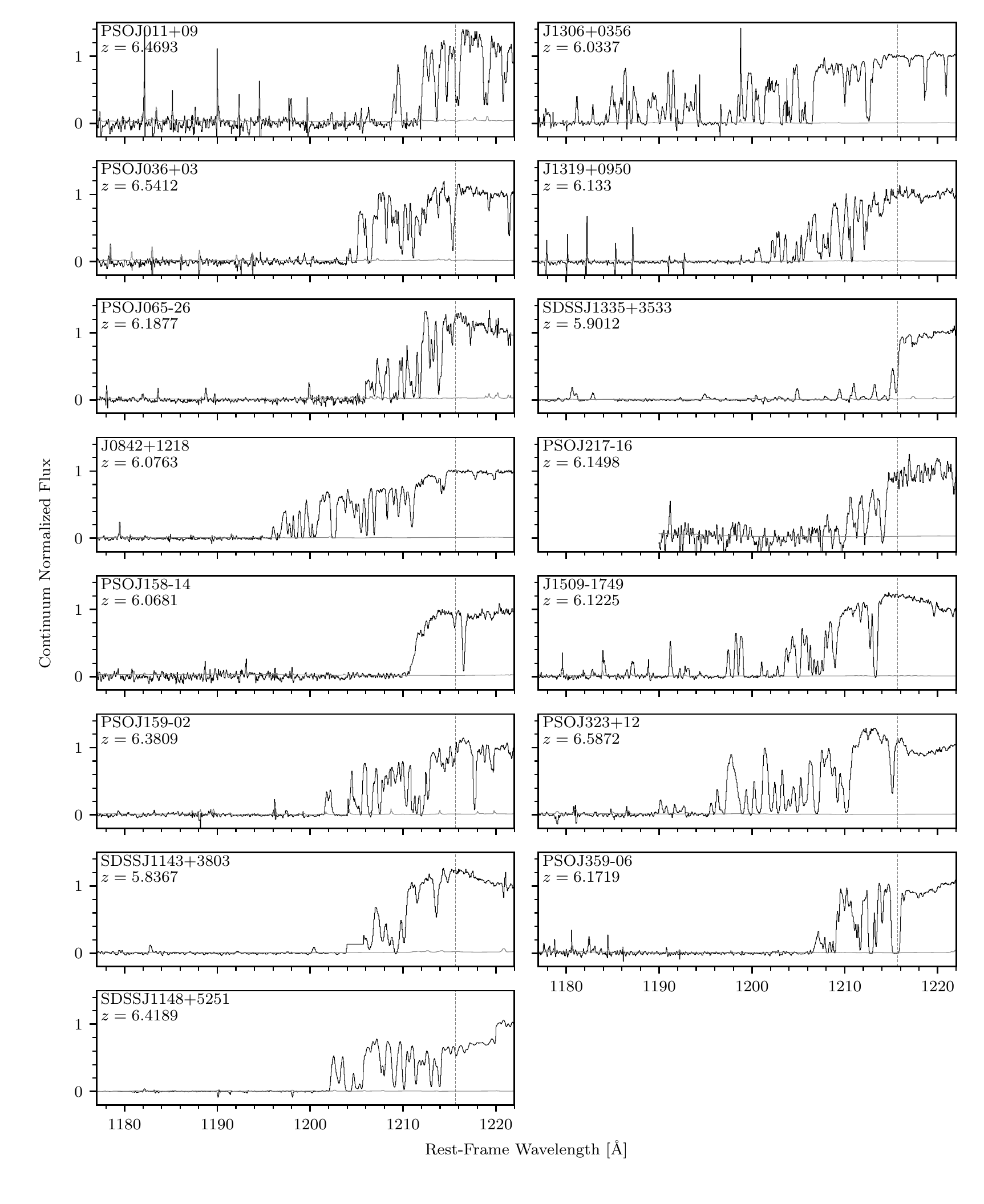}
    \caption{Continuum normalized spectra (black) along with continuum normalized noise vectors (grey) within our sample. The continuum normalized spectra are shown between 1177 and 1222 {\AA}, except for PSOJ217-16, where we only show between 1190 and 1222 {\AA}, due to detector noise. All spectra have been inverse variance smoothed with a 3 pixel filter. The grey dashed line marks the \lya line at 1215.67 {\AA}.}
    \label{fig:cont}
\end{figure*}

To estimate the continuum of the quasar within the proximity zone region, we fit for the quasar continuum redwards of \lya, as this side is unaffected by IGM absorption. We then use this red-side estimate to predict the blue-side continuum, which is affected by absorption \citep{Suzuki2005PredictingForest, Paris2011A3, Davies2018, Bosman2021}.

As a first step to estimate the quasar continua we normalize the matched spectra to unity at $1290\pm2.5$ {\AA}, for ease of fitting.
Then, we sigma clip outlying values at the $3\sigma$ level from the smoothed spectra (10 pixel filter), shown in Figures \ref{fig:spectra1} and \ref{fig:spectra2}, using the noise vectors from each spectra as an estimate of the $\sigma$.
To estimate the quasar continua, we make use of the principal component analysis (PCA) method developed by \cite{Davies2018}, where we use 10 red-side basis spectra (spanning between 1200 and 2850{\AA} for most quasars in the sample) and 6 blue-side basis spectra (spanning between 1177 and 1280 {\AA}). For SDSSJ1143, SDSSJ1148 and SDSSJ1335 we use 10 truncated red-side basis spectra, instead spanning between 1200 and 1450 {\AA} to accommodate these quasars where we do not have NIR coverage.

The principal components were determined on a sample of $12,764$ low-redshift quasars at $z\approx 2$, for which the blue-side continuum is easier to estimate due to the lower level of absorption by the intergalactic medium. \cite{Davies2018} noticed a high degree of correlation or anti-correlation between the red-side coefficients and blue-side coefficients, allowing the blue side coefficients to be estimated from the red side. This can be done using the relationship $b_i = \sum_{j=1}^{10} r_j X_{ji}$, where the $b_i$ and $r_j$ are blue side and red side PCA coefficients, respectively, and $X_{ji}$ is a projection matrix determined by the correlations or anti-correlations between the blue and red side coefficients from the training sample.

Using the MCMC affine-invariant ensemble sampler \verb|emcee| \citep{Foreman-Mackey2013}, we fit each spectrum using $10$ basis spectra redwards of the \lya\ emission line spanning between 1220 {\AA} and 2800 {\AA} in the rest frame, as well as a redshift offset $\delta z$ between the systemic redshift of the quasar and the best redshift estimate for the PCA.
We first take the median of the posterior distribution to estimate the median red side coefficients, we then use the projection matrix $X_{ji}$ to estimate the coefficients for the blue-side continuum. However, unlike \cite{Davies2018} who developed this machinery for studying IGM damping wings in $z > 7$ quasar spectra \citep{Davies2018b}, we match the blue and red side basis spectra at $1220$~{\AA} instead of at $1280$~{\AA} \citep{Eilers2020}. This is due to the fact that all quasars in our sample are at $z \lesssim 6.6$, and thus we do not expect significant absorption redwards of the \lya\ emission line due to IGM damping wings. We set a flat prior on the PCA components to be the $3 \sigma$ boundaries on each of the components determined by \cite{Davies2018}, 
and allow a prior $\delta z \in [-0.01, 0.03]$ for all quasars except for PSO\,J323+12, where we extend the prior to $\delta z \in [-0.05, 0.05]$ 
in order to allow for a better continuum fit. The quasar spectra and their estimated continua are shown in Fig.~\ref{fig:spectra1} and \ref{fig:spectra2}. The continuum normalized spectra that are obtained when dividing by the estimated continuum emission are shown for each quasar in Fig.~\ref{fig:cont}. 


The quasar continuum normalization procedure introduces additional uncertainty into our analysis procedure due to the fact that the continuum estimation can either be an overestimate or underestimate of the true continuum. It can be seen that the PCA based method  underestimates the continuum for SDSSJ1143+3803 and CFHQSJ1509-1749 since the continuum normalized flux is clearly above a flux of unity close to the \lya~line. However, the estimated bias in the PCA continuum resconstruction is very small, i.e. $<1\%$ \citep{Davies2018, Bosman2021} and thus the this underestimation is balanced by the roughly symmetric overestimation of other quasar continua, which can be less obviously identified as it decreases the continuum normalized flux near Ly$\alpha$. We address this uncertainty in the quasar continuum estimates in detail in section \ref{sec:stack}.

\subsection{Radiative Transfer Simulations}

For each quasar in our sample we use
a 1-d radiative transfer (RT) code \citep{Davies2016QuasarMedium} to simulate the effect of ionizing radiation emitted by the quasar along the line-of-sight \citep{Bolton2007AMedium}. We take $1000$ skewers originating from the most massive dark matter halos with halo masses of $4\times 10^{11}\,M_{\odot}\lesssim M_{\rm halo}\lesssim 3\times 10^{12}\,M_{\odot}$ from the cosmological hydrodynamical simulation Nyx with a box of $100\,{\rm Mpc}\,h^{-1}$ on a side \citep{Almgren2013Nyx:COSMOLOGY, Lukic2014TheSimulations}. The RT code computes the time--dependent ionization and recombination rates of six particle species ($e^{-}$, \ion{H}{1}, \ion{H}{2}, \ion{He}{1}, \ion{He}{2}, \ion{He}{3}), as well as the heating and cooling of the IGM due to the expansion of the universe and inverse Compton cooling off the cosmic microwave background. 

The Nyx simulation has outputs at $z= 5.5$, $6.0$ and $6.5$. We choose the output closest to the quasar's emission redshift $z_{\rm em}$ and re-scale the physical densities by $(1+z_{\rm em})^3$. We used the \citet{Lusso2015TheWFC3} composite composite to relate $M_{1450}$ to the specific flux at the Lyman limit, and then compute the ionizing flux by assuming a spectral slope of $\alpha_\nu = -1.70$. 
Previous studies of quasar lifetimes based on proximity zones have only considered the extent of the proximity zone (with the notable exception of \cite{Davies2018b, Davies2019b}), which is defined as the location at which the continuum-normalized flux smoothed to a resolution of $20$~{\AA} in the observed frame drops below the 10\% flux transmission level \citep{Fan2006Quasars}.
Therefore, the precise ionization rate of the ultraviolet background (UVB) did not play a crucial role, since the total ionization rate was always dominated by the quasar at the 10\% flux transmission level. In this work, however, we consider the complete flux profile within the vicinity of the quasar, which extends below the 10\% flux transmission level. At these lower flux transmission levels the ionization rate of the UVB is comparable to the ionization rate of the quasar and therefore needs to be considered, although the UVB should not matter too much given that we are assuming uniformity. Nonetheless, we re-scale the ionization rate of the UVB radiation $\Gamma_{\rm UVB}$ to match the expected strength at the quasar's redshift based on \citet{Haardt2012RADIATIVEBACKGROUND}.

\begin{figure*}[!t]
    \centering
    \includegraphics{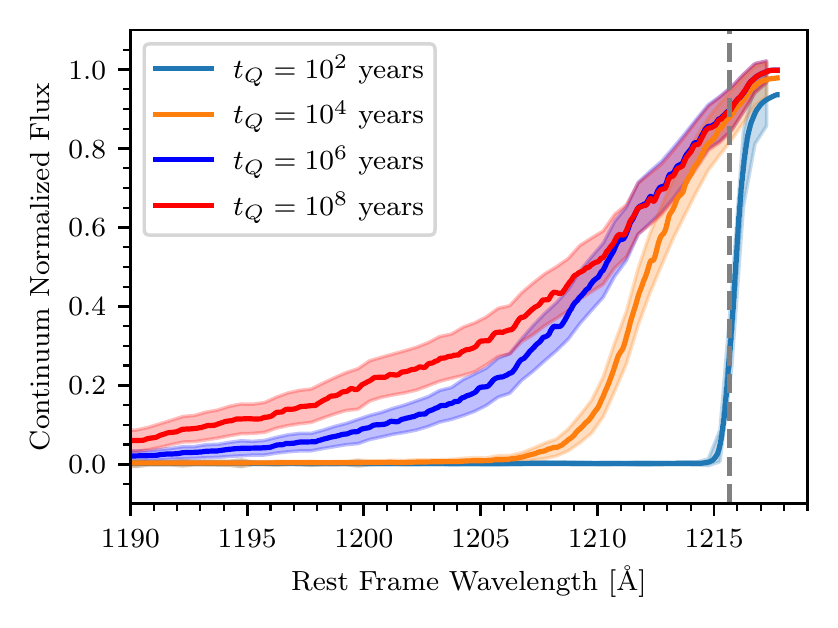}
    \includegraphics{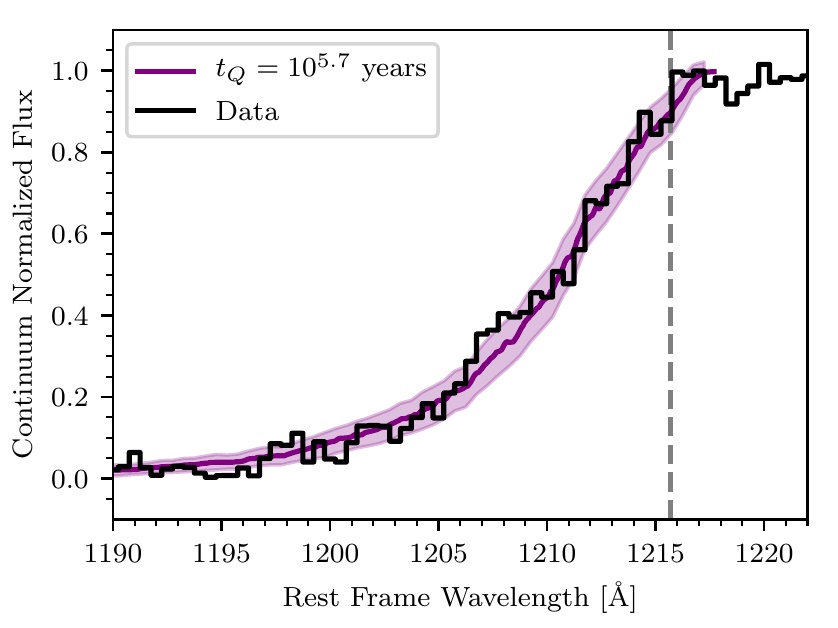}
    \caption{\textit{Left panel:} Proximity zone profile as predicted by radiative transfer simulations for $\log_{10}(t_Q/\rm yr) = $2, 4, 6, and 8. The shaded regions around each of the mean models show the $1\sigma$ uncertainty range as described in section \ref{sec:unc}. The grey dashed line marks the \lya line at 1215.67 {\AA}. 
    \textit{Right panel:} Plot of the quasar stack, with a bin size of 0.5 {\AA}, with the best fit model with $t_Q=10^{5.7}$~yr shown in purple. }
    \label{fig:stack}
\end{figure*}

We then simulate the composite quasar spectrum as a function of quasar lifetime, assuming a light-bulb light curve, ranging from $\log_{10} (t_Q/\rm yr)=1-8.9$ in steps of $\delta \log_{10}(t_Q/\rm yr)=0.1$. Each simulated spectrum is tuned to that quasar's particular redshift and luminosity, and the UVB was calibrated at the particular redshift to match \citet{Haardt2012RADIATIVEBACKGROUND}. 
For each step in the simulation, we compute 1000 different spectra at different sight lines, i.e. we have a model grid of $15 \times 1000 \times 80$ different spectra, since there are 15 different quasars, 1000 different sightlines, and 80 different values of the lifetime, from 1.0 to 8.9 
in $\log t_Q$ space. 
At each time step we stack the simulated spectra by averaging the flux pixels across all 15 quasars and across all 1000 sight lines, since each simulation is computed on the same velocity grid.
The simulated stacked spectra for a few selected lifetimes are shown in the left panel of Fig.~\ref{fig:stack}. 



\section{Results} \label{sec:results}

\subsection{Stack of the Quasars' Proximity Zones}\label{sec:stack}

We now stack the continuum--normalized fluxes within the proximity zone region in order to create a composite flux transmission profile. To this end, we define a regularly spaced grid between 1190 {\AA} and 1218 {\AA} with a grid spacing of $0.5$~{\AA}. We then bin the wavelengths and the flux pixels within this new wavelength grid, i.e. we take an average of all the wavelength pixels and flux pixels within a particular wavelength bin to obtain the stacked flux profile. 
The full composite spectrum within the proximity zone region from the $15$ quasars in our data sample is shown in the right panel of Fig.~\ref{fig:stack}.

\subsection{Forward-Modeling of Uncertainties}\label{sec:unc}


There are two possibilities to estimate uncertainties on the flux pixels within the stacked proximity zone, i.e. either via bootstrapping of the continuum-normalized spectra, or via forward modeling
the various sources of uncertainty onto the simulated quasar spectra. Forward modeling the sources of uncertainty does not give us an estimate of the covariance directly from the data, rather we use covariance estimated from the forward models in performing inference of the quasar lifetime $t_Q$. This gives a better estimate of the covariance than estimating it directly from the data, with the only drawback that the uncertainty on the stack is now model-dependent. 
Due to the limited size of our data sample, the covariance matrix obtained via bootstrapping is very noisy, and thus we forward--model the uncertainties onto the simulated quasar spectra from the RT simulations by means of forward modeling. 
We then estimate the covariance between the different flux bins on the sample of $10,000$ mock stacks with uncertainties. For each stack, we incorporated four types of uncertainty: cosmic variance, uncertainty from the PCA continuum estimation, uncertainty in the systemic redshift, and spectral noise. We assume that the full uncertainties on the stack can be approximated as a multivariate Gaussian distribution about a mean model. 
We will now describe in detail how these uncertainties are being modeled in the quasar stack. We repeat this uncertainty analysis for each lifetime, so we obtain $80$ different samples of $10,000$ mock spectra. We then assume that these mock stacks are from a Gaussian distribution and then estimate the covariance of this distribution directly using the mock stacks.

The first source of uncertainty that we incorporated is uncertainty from cosmic variance. Thus, for each quasar in the stack we randomly chose one of the $1000$ sightlines through the cosmological simulation box, ensuring that every simulated stack consists of $15$ different sightlines. In order to mimic the spectral resolution of the data, we convolve each spectrum with a Gaussian kernel with a standard deviation in velocity space given by $\sigma_\text{vel} = \frac{c}{2.355R}$, where $c$ is the speed of light, $R$ is the resolution of the spectrograph, and the factor of $2.355$ arises from the full width at half maximum (FWHM) to standard deviation conversion. 
Following this, we interpolate the convolved spectrum onto the wavelength grid of the data for the specific quasar in question. That is, we aim to generate simulated data for each quasar that has the same wavelength grid as the actual data, that way we can use the noise from the data to simulate spectral noise.

The \cite{Davies2018} procedure produces a covariance matrix for the continuum prediction, which we can directly sample from to create mock realizations of the uncertainty in the continuum estimate. We then scale the continuum-normalized simulated spectra for each quasar by a draw from this distribution. 
In order to incorporate uncertainties in the systemic redshift estimate of the quasars, we assume an uncertainty on the systemic redshift of $\Delta v=100\,\rm km\,s^{-1}$, and randomly draw from a normal distribution with a mean of $\mu=0$ and the corresponding standard deviation \citep{Eilers2020}. 
The simulated quasar spectra are then shifted to the rest-frame wavelength grid based on this new redshift estimate. Finally, in order to incorporate spectral noise, we directly sample the noise vectors from the data, as we have already interpolated the simulated convolved spectra onto the wavelength grid. To be clear, we are generating sets of 15 quasars each on the wavelength grid of the data, allowing us to directly sample the noise without re-scaling, after which we can stack the 15 quasars within a set using our binning algorithm.



\begin{figure}[t!]
    \centering
    \includegraphics{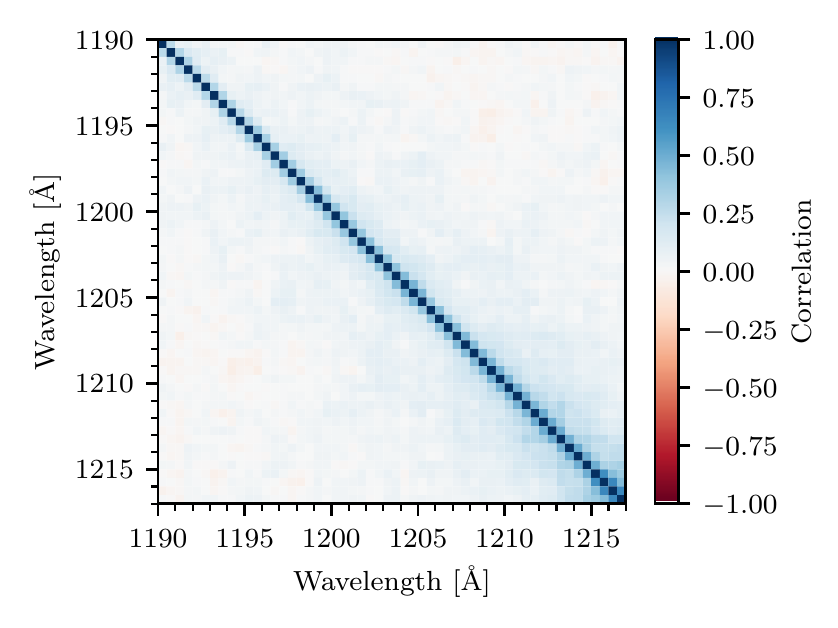}
    \caption{Correlation matrix for $\log_{10}(t_Q/{\rm yr}) = 5.7$ computed from Monte-Carlo sampling the distribution of quasar stacks directly from the simulated quasar spectra from radiative transfer models. }
    \label{fig:corr_model}
\end{figure}

Because the radiative transfer models are dependent on the quasar lifetime, as an input parameter, we performed this Monte-Carlo sampling of mock stacks for each value of the lifetime on a grid between 1.0 and 8.9 in $\log_{10}$ years, with a grid spacing of $\Delta \log_{10}t_Q=0.1$. 
This means that we computed 80 different covariance matrices, for each lifetime dependent model, which we used when we constructed our likelihood function over the log lifetime grid. For each lifetime dependent model we use the same set of random draws for the Monte Carlo sampling to ensure that the likelihood function varies smoothly with the model parameter, rather than being influenced by the stochasticity of the draws.
We show an example correlation matrix for the model corresponding to $t_Q=10^{5.7}$~yr in Fig.~\ref{fig:corr_model}.

\subsection{Quasar Lifetime Estimate} \label{sec:lifetime}

We assume that the data is drawn from a distribution described by one of the MC estimated covariance matrices, centered at the mean model for that particular lifetime. Additionally, we also assume that each quasar in the stack has the same lifetime, i.e. our models are made up of stacks of quasars of the same lifetime (see \S~\ref{sec:summary} for a discussion of this assumption). 

Given our assumptions, we construct a lifetime dependent likelihood function
\begin{equation}
\begin{split}
 \mathscr{L}(r_\lambda|t_Q) =& (2\pi)^{-\frac{k}{2}} \text{det}(\Sigma(t_Q))^{-1/2}  \cdot\exp\bigg(-\frac{1}{2}(f_{\lambda, \text{data}} - f_{\lambda, \text{model}}(t_Q))^{T} \Sigma(t_Q)^{-1}(f_{\lambda, \text{data}} - f_{\lambda, \text{model}}(t_Q))\bigg)   ,
\end{split}
\label{eq:like}
\end{equation}
where $\mathscr{L}(r_\lambda|t_Q)$ is the likelihood of observing the data given a particular model $t_Q$, $k$ is the number of wavelength bins in the stacked spectrum, 
and $\Sigma(t_Q)$ is the model dependent covariance matrix. However, we only know $r_\lambda(t_Q)$ and $\Sigma(t_Q)$ at discrete points in log-lifetime space, since we have outputs of the RT simulation only at discrete time steps. We calculated the likelihood function at the discrete points within the $\log_{10}(t_Q)$ space (i.e. where our covariance matrices are defined) and interpolated the likelihood function to estimate the likelihood between the grid-points.

To construct our likelihood we consider wavelengths from $1190$ to $1217$ {\AA} in the rest-frame. This cutoff at $1190$~{\AA} avoids potential biases in the very low flux regime, where the flux transmission is completely dominated by the UVB. 

\begin{figure}[t!]
    \centering
    \includegraphics[width = \columnwidth]{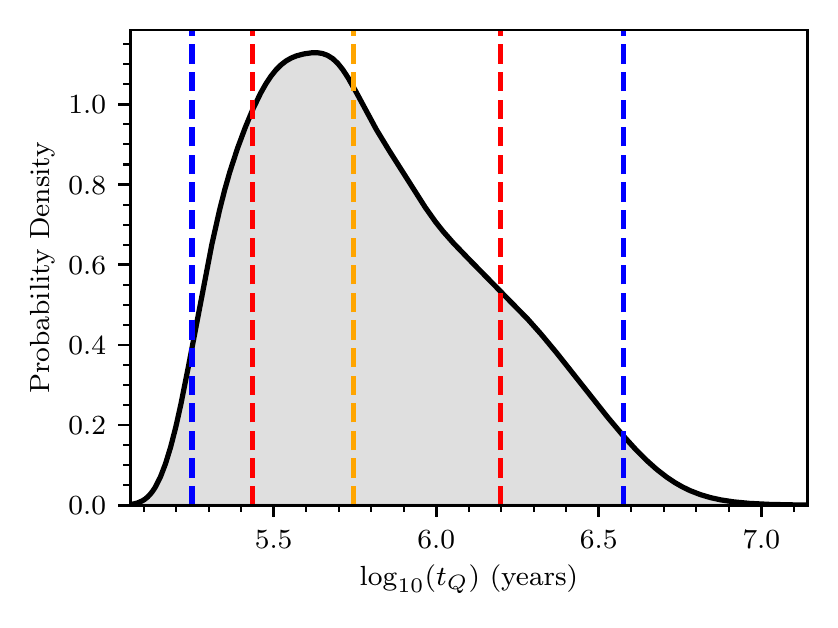}
    \caption{Posterior PDF of the average quasar lifetime $t_Q$. Vertical dashed lines indicate the median (orange), 68\% credible interval (red), and 95\% credible interval (blue).}
    \label{fig:posterior}
\end{figure}

To obtain the posterior probability density function (PDF) for the lifetime, we use a flat prior for $t_Q\in[10, 10^{8.9}]$~yr, meaning that the posterior is merely the normalized version of the likelihood. We normalize the likelihood function using trapezoidal rule numerical integration to obtain the result shown in Fig.~\ref{fig:posterior}. We obtain a measurement of the effective lifetime of the quasar population of $\log_{10}(t_Q/{\rm yr}) = 5.7^{+0.5 (+0.8)}_{-0.3 (-0.5)}$ from the median and 68th (95th) percentile of the posterior probability distribution. This best fit model is shown in the right panel of Fig.~\ref{fig:stack}. 



\begin{figure}[t!]
    \centering
    \includegraphics[width = \columnwidth]{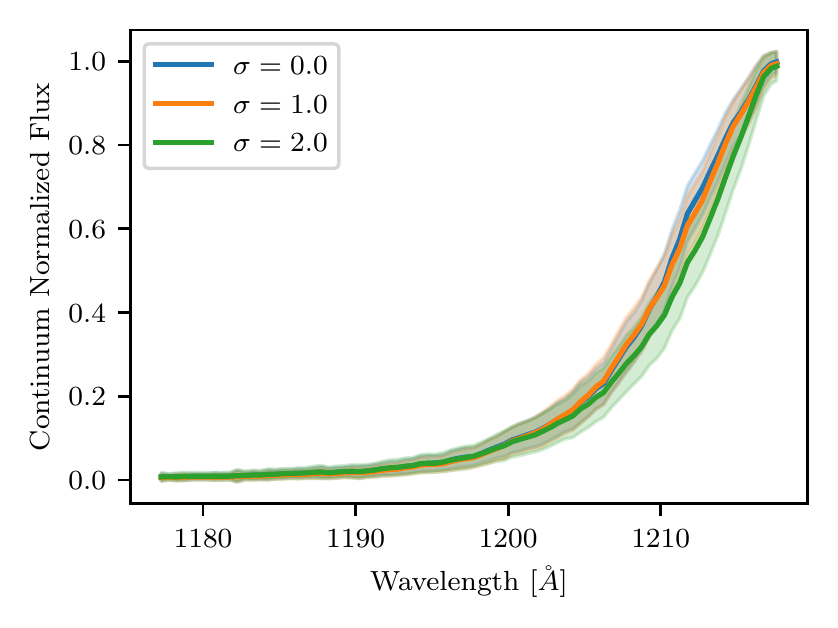}
    \caption{Modeled stacks of the 15 quasars in the stack for a mean lifetime of $\log_{10}(t_Q/{\rm yr})= 5.7$, but for quasar lifetime distribution widths $\sigma = 0, 1, 2$, along with $1\sigma$ uncertainties on the stack.}
    \label{fig:sigmas}
\end{figure}

\section{Summary \& Discussion} \label{sec:summary}

In this paper we estimate the 
effective lifetime of the quasar population at $z\sim6$ by means of the average flux transmission profile in the proximity zone region around the quasars. To this end, we stack the continuum-normalized spectra of $15$ quasars at $5.8 < z < 6.6$ with similar absolute luminosities, i.e. $-27.4 < M_{1450} < -26.6$. For all objects in this data set precise systemic redshift estimates from sub-mm emission lines are available, since uncertainties in the systemic redshifts contribute the largest source of uncertainty to proximity zone estimates \citep{Eilers2017a}. This average composite transmission profile is then compared to simulated transmission profiles from 1-d RT simulations at different effective quasar lifetimes. 

Our best estimate for the effective lifetime of the quasar population as a whole 
based on the composite proximity zone profile is $\log_{10} (t_Q/{\rm yr}) = 5.7 ^{+0.5 (+0.8)} _{-0.3 (-0.5)}$, derived from the median of the posterior probability distribution as well as the $68$th ($95$th) percentile. In order to estimate the uncertainties on the stacked spectrum we forward-model the uncertainties arising due to cosmic variance, the PCA continuum-normalization, the estimate of the quasars' systemic redshifts, as well as spectral noise onto the simulated quasar sightlines. 

Our measurement of the effective lifetime for the quasar population is consistent with previous work, which had estimated an average quasar lifetime of $t_Q\sim 10^6$~yr \citep{Khrykin2019EvidenceQuasars, Davies2019b}. Our results are also consistent with the observed fraction of young quasars within the early universe with lifetimes of $t_Q\lesssim 10^4-10^5$~yr, which was determined to be $5\%\lesssim f_\text{young} \lesssim 10\%$ \citep{Eilers2020, Eilers2021}. This is because we expect to see approximately 10\% of quasars to shine for 10\% of the lifetime, if the quasars' ages are uniformly sampled from $[0, t_Q]$. 

In reality it is likely that the lifetime of the quasar population can not be described by a single value, but rather by a distribution of lifetimes with a mean of $\mu_{t_Q}=t_Q$ and standard deviation $\sigma_{t_Q}$. 
Because we do not measure, or even consider the full distribution of quasar lifetimes, we note that our measurement of the effective lifetime will be comparable to the average lifetime only in the regime where the variance of the quasar lifetime distribution is small. \cite{Khrykin2016TheQuasars} showed that a large variance of the quasar lifetime distribution causes an underestimate of the lifetime, because quasars with lifetime less than $t_Q$ impact the proximity zone more. Additionally, if we have a very young quasar in our stack, this could cause our estimate to be an underestimate of the lifetime.
This is due to the fact that the intergalactic gas reaches ionization equilibrium with the quasar's radiation for lifetimes longer than the equilibration timescale, i.e. $t_Q>t_{\rm eq}\approx3\times10^4$~yr. This means that our measurement of the effective lifetime of the quasar population could potentially  underestimate the mean quasar lifetime, if the width of the distribution is very wide. 
In Figure~\ref{fig:sigmas} we show modeled stacks of the same mean lifetime ($\log_{10}(t_Q/\rm yr) = 5.7$) for the overall  
log-normal distribution of quasar lifetimes, but with different standard deviations. As we can see our method is not really sensitive to the width of the quasar lifetime distribution, unless the width of the lifetime distribution is large, i.e. $\sigma_{\log_{10}t_Q}\gtrsim 2$, in which case our method would underestimate the average quasar lifetime. 
However, recent results suggest only a very modest standard deviation of the distribution of quasar lifetimes, i.e. $\sigma_{\log_{10}t_{\rm Q}}=0.80^{+0.37}_{-0.27}$ \citep{Khrykin2021}, suggesting that our measurement of the effective quasar lifetime actually reflects the average lifetime of the quasar population as a whole. 


Another potential source of uncertainty in our measurement is the systematic bias in the PCA continuum estimation. \cite{Davies2018} estimate that the PCA reconstructed continuum overestimates the true quasar continuum on average by $\sim 1\%$. However, this is a relative error on the flux, i.e. at $\sim 10\%$ flux transmission the estimated error is $0.1\%$ and hence this bias has only negligible effects on our final results. 

Our lifetime measurement is more than an order of magnitude shorter than expected from exponential growth models for SMBHs \citep{Volonteri2010FormationHoles, Volonteri2012TheHoles}. 
A potential explanation for such short average quasar lifetimes could be flickering quasar light curves instead of the simple ``light-bulb'' light curves, implying that there could be multiple epochs of quasar activity and concurrent black hole growth. Since the transmitted flux within the proximity zone around quasars at $z\sim 6$ is only sensitive to the last ``on-time'' our composite proximity zone spectrum from this analysis is not sensitive to potential previous phases of quasar activity. 

However, several analysis have shown previously that even flickering quasar light curves cannot explain the discrepancy between the short average quasar lifetimes and the time required to explain the SMBH growth with the current exponential growth model. For instance, \citet{Davies2019a} showed that the damping wing feature observed in quasar spectra at $z>7$, where the surrounding IGM still has a very high neutral gas fraction, provides a constraint on the \emph{total} 
number of ionizing UV photons emitted into the IGM integrated over cosmic time, irrespective of the light-curve shape. 
Their analysis shows that the integrated
time these quasars have been emitting UV
radiation is still consistent with a light bulb light-curve with
$t_Q\sim10^6$ years. 

At $z\sim6$, however, the IGM is highly ionized and the quasar spectra do not exhibit a damping wing. Hence the proximity zones are sensitive only to the on-time of the last accretion episode, which is only consistent with the total quasar lifetime for light-bulb light curves. \citep{Davies2019b} has analyzed the distribution of $z\sim 6$ proximity zones in the context of various quasar light curves. While they find a good agreement between the observed distribution of proximity zones and model predictions when assuming a simple light-bulb light curve, flickering quasar light curves with multiple black hole growth phases cannot be excluded for $z\sim6$ quasars. 


Our presented results at $z\sim 6$, the distribution of proximity zones at $z\sim 6$ \citep{Davies2019b}, the fraction of very young quasars within the quasar population \citep{Eilers2020, Eilers2021}, as well as the damping wing analysis at $z>7$ \citet{Davies2019a}, are all in good agreement and point towards quasar lifetimes of $\log_{10}(t_Q) \simeq 10^6$. These measurements suggests that either a significant fraction of the black hole growth occurs in UV obscured, dust-enshrouded growth phases in the early universe, or it could indicate radiatively inefficient mass accretion rates, such that more of the accreted mass contributes to the growth of the black holes rather than the quasars' UV emission \citep{Eilers2018b, Davies2019a}. In future work, we aim to explore possible mechanisms by which quasars can accrete sufficient mass while only emitting UV radiation for $\sim 10^6$ years on average. Additionally, we will expand this work to measure the properties of the full quasar lifetime distribution from the individual measurements, as suggested in \citep{Khrykin2021}. These studies will provide further insights for our understanding of SMBH formation in the early universe and whether or how the accretion mechanisms change with cosmic time.

\acknowledgements
The authors would like to thank the UROP office at MIT for generous funding of this work. \\

ACE acknowledges support by NASA through the NASA Hubble Fellowship grant $\#$HF2-51434 awarded by the Space Telescope Science Institute, which is operated by the Association of Universities for Research in Astronomy, Inc., for NASA, under contract NAS5-26555. 

\bibliography{references2}

\end{document}